\documentclass[notitlepage]{article}
\usepackage{amssymb}

\newtheorem{remark}{Remark}[section]
\newtheorem{theorem}{Theorem}[section]
\newtheorem{corollary}{Corollary}[section]
\newtheorem{definition}{Definition}[section]
\newtheorem{conjecture}{Conjecture}[section]
\newtheorem{assumption}{Assumption}[section]
\newtheorem{lemma}{Lemma}[section]

\newcommand{\rank}{\mathrm{rank\;}}
\newcommand{\ddim}{\mathrm{ddim\;}}
\newcommand{\dind}{\mathrm{dind\;}}

\newcommand{\sgrad}{\mathrm{sgrad\;}}
\newcommand{\reg}{\mathrm{reg\;}}
\newcommand{\ann}{\mathrm{ann\;}}
\newcommand{\Span}{\mathrm{span\;}}

\begin{document}

\pagestyle{plain}

\title{Non-commutative Integrability,
Moment Map and Geodesic Flows}

\author{
Alexey V. Bolsinov \\
{\small Moscow State University}\\ 
{\small Department of Mechanics and Mathematics, 119899, Moscow, Russia} \\
{\small e-mail:  {\tt bolsinov$@$mech.math.msu.su}}\\
\and
Bo\v zidar Jovanovi\' c \\
{\small Matemati\v cki Institut, SANU}\\
{\small Kneza Mihaila 35, 11000 Beograd, Serbia, Yugoslavia} \\
{\small e-mail: {\tt bozaj$@$mi.sanu.ac.yu}}
}
\date{}

\maketitle

\begin{abstract}
The purpose of this paper is to discuss the relationship between
commutative and non-commutative integrability  of Hamiltonian systems
and to construct new examples of integrable geodesic flows on
Riemannian manifolds. In particular, we prove that the geodesic flow
of the bi-invariant metric on any bi-quotient of a compact Lie group is
integrable in the non-commutative sense by means of polynomial integrals, and
therefore, in the classical commutative sense by means of $C^\infty$--smooth
integrals.
\end{abstract}

{\small

\centerline

\medskip

\centerline{{\bf Key words:} integrable Hamiltonian systems,
geodesic flows,} 
\centerline{Hamiltonian action of a Lie group, non-commutative integrability}
}

\section{Introduction}

The purpose of this paper is to establish complete integrability of
a wide class of Hamiltonian systems connected with Hamiltonian
actions of Lie groups,
with a special attention to the integrability of the geodesic
flows of Riemannian metrics.

Let $(M,\omega)$ be $2n$--dimensional connected symplectic manifold.
By $\sgrad h$
we shall denote the Hamiltonian vector field of a function $h$,
and by $\{\cdot,\cdot\}$ the canonical Poisson brackets
on $M$. Consider Hamiltonian equations:
$$
\dot x=\sgrad h(x).
\leqno(0.1)$$

A function $f$ is an integral of the Hamiltonian system  (constant
along trajectories of (0.1)) if and only if it
commutes with $h$: $\{h,f\}=0$.

One of the central problems in Hamiltonian dynamics is
whether the equations (0.1) are completely integrable or not.
The usual definition of complete integrability is as follows
(see for instance [20]):

\begin{definition} {\rm
Hamiltonian equations (0.1) are called {\it completely integrable}
if  there are
$n$ Poisson-commuting smooth integrals $f_1,\dots,f_n$ whose
differentials are independent in an open dense subset of $M$.}
\end{definition}

If system (0.1) is completely integrable then by Liouville's theorem
the general solutions of the system (0.1) can be (locally)
solved in quadratures.
Moreover, compact connected components of regular invariant submanifolds
$\{f_1=c_1,\dots,f_n=c_n\}$
are diffeomorphic to $n$--dimensional Lagrangian tori
with linear dynamics (see [1, 2, 18, 20, 33]).

The algebra of integrals $\mathcal F=\{f_1,\dots,f_n\}$ is called
a {\it  complete involutive (or commutative) algebra} of functions on $M$.

Let $M$ be the cotangent bundle of a Riemannian manifold
$(Q,g)$ with the natural symplectic structure.
Taking  $h(\xi,x)=\frac12 g^{-1}_x(\xi, \xi)$, $\xi\in
T^*_xQ$ as a Hamiltonian in (0.1), we obtain the equations of
the geodesic flow.  It is very rare for this flow to be completely
integrable.  Almost all known manifolds with integrable geodesic flows are
diffeomorphic to some symmetric spaces. The point is that there are very
serious topological obstructions to the integrability. As an example, we
mention Taimanov's theorem [30, 31] saying that in the real
analytic case the fundamental group a manifold $Q$ admitting integrable
geodesic flows must be almost commutative ($Q$ has a finite-sheeted
covering $p: \tilde Q\to Q$, where the group $\pi_1(\tilde Q)$ is commutative).

The classical examples of Riemannian manifolds with integrable geodesic
flows are surfaces of revolution (Clairaut), flat tori, $n$--dimensi\-onal
ellipsoids (Jacobi) and the Lie group $SO(3)$ with a left invariant metric
(Euler).  Mishchenko and Fomenko proved integrability
of certain left-invariant metrics on compact Lie groups
[22, 33]. This result was generalized to all compact symmetric spaces
by Mishchenko [21], Thimm [32], Mikityuk [24] and Brailov [10]. Examples of
homogeneous, but not symmetric, spaces with integrable geodesic flows can
be found in [6, 7, 11, 16, 25, 28, 32].
Recently the authors [8] have proved the non-commutative integrability
of geodesic flows on all homogeneous spaces
$\mathfrak G/\mathfrak H$, where $\mathfrak G$ is a compact connected Lie group.

Natural generalizations of homogeneous spaces are the so-called
bi-quotients of Lie groups.
These manifolds were introduced by Gromoll and Meyer [15]
in order to construct metrics with nonnegative sectional curvature
on  the exotic 7--sphere.

The examples of bi-quotients with integrable geodesic flows
(which include, in particular, Eschenburg's manifolds [14] and the exotic
7--sphere of Gromoll and Meyer [15]) have been found and studied
by Paternain and Spatzier [28] and Bazaikin [3].

Now we shall present the main results of the paper:

\begin{itemize}

\item We prove that non-commutative integrability 
always implies complete commutative
integrability in the sense of definition 0.1 by means of
$C^\infty$--smooth integrals (section 1). This result proves the
conjecture of Mishche\-nko and Fomenko [33] about the relationship
between these two kinds of integrability in the $C^\infty$--case.

\item Consider the Hamiltonian action of a Lie group $\mathfrak G$
on a symplectic manifold $M$ with a moment map $\Phi:M\to\mathfrak g^*$.
Under some assumption (that holds if the action is proper) we prove that from
$\mathfrak G$ invariant functions and so-called collective functions
(functions of the form: $f_h=h\circ\Phi$) we can always construct 
completely integrable systems on $M$. In that way we generalize
the results of Guillemin and Sternberg [16, 17, 18] obtained for
multiplicity-free actions (section 2).

\item As a simple application of the above results, we prove
the complete integrability of geodesic flows on manifolds all
of whose geodesics are closed (section 3) and
give a new prove of the theorem that
geodesic flows on homogeneous spaces $\mathfrak G/\mathfrak H$
of compact Lie groups $\mathfrak G$
are completely integrable (section 4).

\item Finally, we prove integrability of geodesic flows
of bi-invariant metrics
on all bi-quotients
$\mathfrak K\backslash \mathfrak G/\mathfrak H$ of compact Lie groups $\mathfrak G$
(section 5).

\end{itemize}

The notation of the paper is usual.
The algebra of smooth functions on a manifold $M$
is  denoted by $C^\infty(M)$.
If something holds {\it for general} $x\in M$, this means
that the property holds for all $x$ from an open dense
subset of $M$.
By $\mathfrak G,\mathfrak H,\mathfrak K,\dots$ we denote
Lie groups and by $\mathfrak g,\mathfrak h,\mathfrak k,\dots$
the corresponding Lie algebras.
The orbit of the co-adjoint action through $\mu\in\mathfrak g^*$
is denoted by ${\mathcal O}(\mu)$.
For a $\mathfrak G$ action on $M$, the orbit through $x\in M$
is denoted by $\mathfrak G \cdot x$ and
the isotrop subgroup of $x$ by $\mathfrak G_x$.
The annihilator of the linear subspace $V\subset W$
in the dual space $W^*$ is denoted by
$\ann (V)\subset W^*$.

\subsection*{Acknowledgments}
First of all we are very grateful to Prof. A. S. Mish\-chenko for the
extremely useful discussion which helped us to look at the problem
from a more general point of view and, as a result, to simplify
proofs and generalize some results. 
This paper was written while the first
author was a guest professor at Freiburg University
and the second author was on a postdoc position at the
Graduiertenkolleg "Mathematik in Bereich ihrer Wechselwirkung mit der Physik",
of the Mathematisches Institut, LMU, M\" unchen.
We would like to use this opportunity to thank
Albert-Ludwigs-Universit\"at and Ludwig Maximillians
Universit\"at for their hospitality.
The first author was also supported by Russian Found for Fundamental
Research (grants 99-01-01249 and 00-15-99272).
The second author was partially supported by the Serbian Ministry
of Science and Technology, Project 1643 (Geometry and Topology
of Manifolds and Integrable Dynamical Systems).

\section{Non-commutative integrability}

We shall briefly recall the concept of non-commutative integrability introduced by Mishchenko and Fomenko in [23].

If $f_1$ and $f_2$ are integrals of (0.1), then so are an
arbitrary smooth function $F(f_1,f_2)$ and the Poisson bracket $\{f_1,f_2\}$.
Therefore without loss of generality we can assume that integrals of (0.1)
form an algebra $\mathcal F$ with respect to the Poisson bracket.
For simplicity we shall assume that $\mathcal F$ is functionally generated by
functions $f_1,\dots,f_l$ so that:
$$
\{f_i,f_j\}=a_{ij}(f_1,\dots,f_l).
\leqno(1.1)
$$

Suppose that
\begin{eqnarray*}
&&\dim F_x=\dim \Span\{df_i(x),i=1,\dots,l\}=l, \quad x\in U, \\
&&\dim \ker\{\cdot,\cdot\}\vert_{F_x}=r, \quad x\in U,
\end{eqnarray*}
holds for an open dense set $U\subset M$.

Let $\phi: M\to \Bbb{R}^l$ be the moment mapping:
$$
\phi(x)=(f_1(x),\dots,f_l(x))
\leqno(1.2)$$
and let $\Sigma=\phi(M\setminus U)$.

Then we have the following
non-commutative integration theorem (see also Mishchenko and Fomenko [23],
Nekhoroshev [9] and Brailov [26]):

\begin{theorem}Suppose that:
$$
\dim F_x + \dim \ker\{\cdot,\cdot\}\vert_{F_x}=\dim M,
\leqno(1.3)$$
for $x\in U$.
Let $c\in \phi(M)\setminus \Sigma$ be a regular value of the moment map.
Then:

(i) $M_c=\phi^{-1}(c)$ is an isotropic submanifold of $M$ and
the equations (0.1) on $M_c$ can be (locally) solved by quadratures;

(ii) Compact connected components $T^r_c$ of $M_c$ are diffeomorphic to
$r$--dime\-nsional tori.
In the neighborhood of $T^r_c$ there are {\it generalized action-angle
variables} $y,x,I,\varphi\mathrm{mod\;}2\pi$
such that the symplectic form becomes:
$$
\omega=\sum_{i=1}^r dI_i\wedge d\varphi_i+\sum_{i=1}^q dy_i\wedge dx_i,
$$
and the Hamiltonian function $h$ depends only on $I_1,\dots,I_r$.
The invariant tori are given as the level sets of integrals $I_i,y_j,x_k$.
The equations (0.1) on  invariant tori take the linear form:
$$
\dot \varphi_1=\omega_1(I)=\frac{\partial h}{\partial I_1},
\dots, \dot \varphi_r=\omega_r(I)=\frac{\partial h}{\partial I_r}.
$$
\end{theorem}

Let us just point out some steps in the proof of the theorem.

Under the hypotheses of theorem 1.1,  there exist $r$ linearly
independent commuting vector fields $X_1=\sgrad h,X_2,\dots,X_r$ on $M_c$.  
They
can be obtained as linear combinations of the Hamiltonian vector fields
$\sgrad f_1,\dots,\sgrad f_l$ from the conditions $\omega(X_i,\sgrad f_j)=0$,
$i=1,\dots,r$, $j=1,\dots,l$.  Since commutative algebras are solvable we
can apply the theorem of S.Lie to integrate the system $\dot x=\sgrad
h(x)$.

The Lie theorem says that if in some domain $V\subset \Bbb{R}^n\{x\}$,
we have $n$ linearly independent vector fields $X_1,\dots,X_n$ that
generate a solvable Lie algebra under commutation, and if
$[X_1,X_i]=\lambda_i X_i$, then the differential equation
$\dot x=X_1(x)$ can be integrated by quadratures in $V$ (see [2, 34]).
In our case $V$ is an open set in $M_c$, $n=r$.

On a compact connected component $T^r_c$ of $M_c$, the vector fields
$X_1,\dots,X_r$ are complete. Therefore $T^r_c$ is diffeomorphic to
an $r$--dimensional torus and
the motion on the torus is quasi-periodic
(the proof of this fact is just the same
as in the usual Liouville theorem [1]).
The existence of generalized
action-angle variables follows from the Nekhoroshev theorem [26].

If the algebra of integrals $\mathcal F=\{f_1,\dots,f_l\}$ is commutative,
i.e. $\{f_i,f_j\}=0$ and $n=l=r$ then from theorem 1.1 we obtain
the Liouville theorem. But,
in fact theorem 1.1 is essentially stronger then Liouville's
theorem: under the assumptions of theorem 1.1,
the Hamiltonian system (0.1) is integrable in the usual, commutative sense.

In order to prove this statement, we shall modify
a remark by Brailov (see [34]) that on 
every symplectic manifold there exists a complete involutive set
of functions. Brailov's idea was to fill up the manifold with
disjoint Darboux symplectic balls. On every symplectic ball
one can construct complete involutive set of functions that can be then
"glued" in order to obtain involutive functions globally defined.

\begin{theorem}Under the assumptions of theorem 1.1,
the Hamiltonian system (0.1) is integrable in the usual, commutative sense,
i.e., it admits $n$ Poisson-commuting $C^\infty$--smooth
integrals $g_1,\dots,g_n$,
independent on an open dense subset of $M$.
\end{theorem}

{\it Proof.}
On the image of $M$ under the moment mapping
$\phi(M)\subset\Bbb{R}^l\{y_1,\dots,y_l\}$ we can introduce
a "Poisson structure" $\{\cdot,\cdot\}_\mathcal F$ by the formulas:
$$
\{y_i,y_j\}_{\mathcal F}=a_{ij}(y_1,\dots,y_l).
$$
Note that $\rank\{\cdot,\cdot\}_{\mathcal F}(y)=l-r=2q$
for all $y\in \phi(M)\setminus \Sigma$
and that $(\phi(M)\setminus \Sigma,\{\cdot,\cdot\}_{\mathcal F})$  is a
Poisson manifold.

From the definition of the Poisson brackets $\{\cdot,\cdot\}_{\mathcal F}$  it
follows that if smooth functions $F,G: \Bbb{R}^l\to\Bbb{R}$ are in
involution with respect to $\{\cdot,\cdot\}_{\mathcal F}$ on $\phi(M)$, then
their liftings $f=F\circ\phi$ and $g=G\circ\phi$ commute on $M$:
$$
\{f,g\}(x)=\{F,G\}_{\mathcal F} (\phi(x))=0.
\leqno(1.4)
$$

Let $z\in \phi(M)\setminus\Sigma$. By the theorem on the local structure
of Poisson brackets (see [20]), there is a neighborhood
$U(z)\subset \phi(M)\setminus\Sigma$ of $z$ and
$l$ independent smooth functions
$G_1,\dots,G_l: U(z)\to\Bbb{R}$, $G_i(z)=0$ such that
$$
\{G_i,G_{i+q}\}_{\mathcal F}=1=-\{G_{i+q},G_i\}_{\mathcal F}
 \quad i=1,\dots, q
\leqno(1.5)$$
and the remaining Poisson brackets vanish.
Let a  ball $B^\alpha(\epsilon_\alpha)$ belong to $U(z)$, where
$$
B^\alpha(\epsilon_\alpha)=\{y\in U(z), G_1^2+\dots +G_l^2 < \epsilon_\alpha \}
\leqno(1.6)$$
Then, starting from the $n$ involutive functions on $B^\alpha$:
$$
h_1=G_1^2+G_{1+q}^2,\dots,h_q=G_q^2+G_{2q}^2,
h_{q+1}=G_{2q+1}^2, \dots, h_{n}=G_{l}^2
\leqno(1.7)$$
we can construct a smooth set
of nonnegative functions $F^\alpha_1,\dots,F^\alpha_n: \Bbb{R}^l\to\Bbb{R}$
that are independent on an open dense subset of
$B^\alpha(\epsilon)$, equal to zero outside $B^\alpha(\epsilon_\alpha)$,
in involution on $\phi(M)$,
and satisfies inequalities
$F_i^\alpha < e^{-\epsilon_\alpha}$
together with all derivatives.
To this end we use the construction suggested by Brailov for Darboux
symplectic balls (see [34]).

Let $g: \Bbb{R}\to\Bbb{R}$ be smooth nonnegative function,
such that $g(x)$ is equal to zero for $\vert x \vert > \epsilon_\alpha$,
monotonically increases on $[-\epsilon,0]$ 
and monotonically decreases on $[0,\epsilon]$.
Let $h(y)=g(h_1(y)+\dots+h_n(y))$.
This function could be extended
by zero to the whole manifold. Now, we can define $F^\alpha_i$ by:
$F^\alpha_i=h\cdot h_i$. Obviously, $\{F^\alpha_i,F^\alpha_j\}_{\mathcal F}=0$.
These functions are independent inside $B^\alpha$.
Also, we can choose $g$ in such a way that $F^\alpha_i$
satisfies the inequalities $F_i^\alpha < e^{-\epsilon_\alpha}$
together with all derivatives.

In the same way we can construct a countable family of open balls
$\{B^\alpha(\epsilon_\alpha)\}$,
$B^\alpha \cap B^\beta =\emptyset$,
and functions $\{F^\alpha_1,\dots,F^\alpha_l\}$ with
the above properties, such that $B=\cup_\alpha B^\alpha(\epsilon_\alpha)$
is an open everywhere dense set of $\phi(M)\setminus\Sigma$.
Let us define the functions $F_1,\dots,F_n: \Bbb{R}^l\to \Bbb{R}$ by:
\begin{eqnarray*}
&&F_i(y)=F_i^\alpha, \quad y \in B^\alpha\subset B \\
&&F_i(y)=0, \quad y \in \Bbb{R}^l\setminus B,\quad i=1,\dots,n
\end{eqnarray*}

By (1.4), the functions $g_1=F_1\circ\phi,\dots,g_n=F_n\circ\phi$
will have the desired properties.
q.e.d.

\medskip

Theorem 1.2 says that the $r$--dimensional invariant tori $T^r$
can be organized into larger, $n$--dimensional Lagrangian tori $T^n$
that are level sets of a commutative algebra of integrals.
Since the tori $T^n$ are fibered into invariant tori $T^r$,
the trajectories of (0.1) are not  dense on $T^n$. In this sense,
the system (0.1) is degenerate.  Thus, establishing the fact of
non-commutative integrability of the system give us more information on
the behavior of its integral trajectories than we could obtain from the
usual Liouville integrability.  Note that, contrary to the case of
non-degenerate integrable systems, the fibration by Lagrangian tori is
neither intrinsic nor unique.

Let $P_0=\phi(M)\setminus\Sigma$, $M_0=\phi^{-1}(P_0)$.
If all invariant submanifolds
$M_c=\phi^{-1}(c)$, $c\in P_0$ are compact and
connected, then $\phi: (M_0,\{\cdot,\cdot\})\to
(P_0,\{\cdot,\cdot\}_{\mathcal F})$ is a Poisson morphism and isotropic
fibration. This fibration is symplectically complete, i.e., the
symplectic orthogonal distribution to the tangent spaces of
the fibres is
a foliation (see [20]). The geometry of such fibrations as well as
obstructions to the existence of global generalized action-angle variables
are studied in [12, 26].

By theorems 1.1 and 1.2, we can give the following definition of
completely integrable systems:

\begin{definition}{\rm
Let  $\mathcal F \subset C^\infty(M)$ be an algebra of functions,
closed under the Poisson brackets.
Let $F_x$ be the subspace of $T^*_xM$ generated by $df(x)$, $f\in \mathcal F$.
Let $K_x\subset F_x$ be the kernel of Poisson structure restricted on $F_x$.
Suppose that $\dim F_x=l$, $\dim K_x=r$ holds on an open dense subset $U
\subset M$. We shall denote  $U$ by $\reg{\mathcal F}$  ({\it
regular points} of $\mathcal F$).  The numbers $l$ and $r$ are usually denoted
by $\ddim \mathcal F$ and $\dind \mathcal F$ and are called {\it differential
dimension} and {\it differential index} of $\mathcal F$.  The algebra $\mathcal F$
is said to be {\it complete} if:
$$
\ddim{\mathcal F}+\dind{\mathcal F}=\dim M.
$$
}\end{definition}

\begin{remark}{\rm
Let $\mathcal F$ be a complete algebra,  $x\in\reg{\mathcal F}$ and
$$
W_x=\Span \{\sgrad f(x),\; f\in {\mathcal F}\}.
$$
Then the completeness condition
$\dim F_x+\dim \ker \{\cdot,\cdot\}_{F_x}=\dim M$ is
equivalent to the coisotropy of $W_x$ in the symplectic linear space $T_xM$:
$W_x^\omega \subset W_x$.}
\end{remark}

\begin{remark}{\rm
Suppose that independent functions $f_1,\dots,f_l$ generate
a finite dimensional Lie algebra
$\mathfrak g=\mathcal F=\oplus_{i=1}^l\Bbb{R}f_i$
under the Poisson brackets:
$$
\{f_i,f_j\}=\sum_{k=1}^l c^k_{ij}f_k,
$$
$c^k_{ij}$ are constants.
In that case the numbers $\ddim \mathcal F$ and $\dind \mathcal F$ coincide with
the dimension and the index of the Lie algebra $\mathfrak g$ [23].
}
\end{remark}

\begin{definition}{\rm
The Hamiltonian system (0.1) is {\it completely
integrable}
if it possesses a  complete algebra $\mathcal F$ of integrals.}
\end{definition}

\begin{remark}{\rm
In definition 1.2 we do not require that
$\mathcal F$ is
generated by $l=\ddim\mathcal F$ functions.
We shall briefly explain this.
Let $x_0$ belong to $\reg\mathcal F$.
Then there are integrals $f_1,\dots,f_l\in \mathcal F$ that are
independent in $x_0$, where $l=\ddim \mathcal F$.
Let $V$ be the open set where
the functions $f_1,\dots,f_l$ are independent.
Since $f_1,\dots,f_l$ are integrals of (0.1) the trajectory of (0.1)
that has initial position in $V$ remains in $V$. Indeed,
the phase flow of (0.1) preserves the
form $df_1\wedge\dots\wedge df_l$.
Therefore we can consider the restriction of (0.1) to $V$ and apply
theorem 1.1 to integrate it.
If all connected regular invariant submanifolds are
isotropic tori
then by the same construction
as in theorem 1.2, we can construct a commutative set of smooth
first integrals that are independent on an open dense set of $M$.
The original isotropic tori are organized in Lagrangian tori that are level
sets of a commutative algebra of integrals.}
\end{remark}

Mishchenko and Fomenko stated the conjecture that  non-commutative
integrable systems are integrable in the usual commutative sense by means of
integrals that belong to the same functional class as the original
non-commutative algebra of integrals.
In particular,
when $\mathcal F$ is a
finite-dimensional Lie algebra, this conjecture is proved
for compact manifolds $M$ and for all semi-simple Lie
algebras (see [5, 33, 34] and references therein).

Our theorem 1.2 proves the conjecture in general 
without assuming that $\mathcal F$ is finite dimensional.
However the above construction gives the proof only for
$C^\infty$-smooth integrals and does not allow us to get
real analytic ones.
Thus the following
general conjecture remains:

\begin{conjecture}{\rm Suppose that on a real-analytic symplectic
$2n$-dimensional manifold $M$ we have
a Hamiltonian system $\dot x=\sgrad h(x)$ completely integrable by means of
a non-commutative algebra $\mathcal F$ of integrals which are real analytic
functions.  Then the system possesses $n$ commuting real analytic
integrals.}
\end{conjecture}

\section{Moment map. Collective motion}

Let a connected Lie group $\mathfrak G$ act on $2n$--dimensional connected
symplectic manifold $(M,\omega)$.
Suppose the action is Hamiltonian.
This means that $\mathfrak G$ acts on $M$ by  symplectomorphisms
and there is a well-defined momentum mapping:
$$
\Phi: M\to \mathfrak g^*
\leqno(2.1)
$$
($\mathfrak g^*$ is a dual space of the Lie algebra $\mathfrak g$)
such that
one-parameter subgroups of symplectomorphisms  are generated by the
Hamiltonian vector fields of functions
$f_\xi(x)=\Phi(x)(\xi)$, $\xi\in\mathfrak g$, $x\in M$  and
$f_{[\xi_1,\xi_2]}=\{f_{\xi_1},f_{\xi_2}\}$.
Then $\Phi$
is  equivariant with respect to the given action of  $\mathfrak G$ on $M$
and the co-adjoint action of $\mathfrak G$ on $\mathfrak g^*$:
$$
\Phi(g\cdot x)=Ad_g^*(\Phi(x)).
\leqno(2.2)
$$
In particular, if $\mu$ belongs to $\Phi(M)$ then the co-adjoint
orbit ${\mathcal O}(\mu)$ belongs to $\Phi(M)$ as well.

Throughout the paper we shall use the following notation for
two natural classes of functions on $M$.
By ${\mathcal F}_1$ we shall denote the subalgebra of functions in
$C^\infty(M)$ obtained by pulling-back the algebra $C^\infty(\mathfrak g^*)$
by the moment map (2.1):
$$
{\mathcal F}_1=\Phi^*C^\infty(\mathfrak g^*)=\{f_h=h\circ\Phi, \;
h:\mathfrak g^*\to \Bbb{R}\}
\leqno(2.2)
$$
and by
${\mathcal F}_2$ we shall denote the algebra of $\mathfrak G$--invariant
functions in $C^\infty(M)$:
$$
{\mathcal F}_2=\{ f: M \to \Bbb{R}, \; f(g\cdot x)=f(x),
\; x\in M, g\in\mathfrak G \}.
\leqno(2.3)
$$

Let  $\{\cdot,\cdot\}_{\mathfrak g^*}$ be the Lie-Poisson bracket on $\mathfrak g^*$:
$$
\{f(\mu),g(\mu)\}_{\mathfrak g^*}=\mu([df(\mu),dg(\mu)]),
\quad f,g: \mathfrak g^*\to \Bbb{R}.
\leqno(2.4)
$$
Then the mapping $h\mapsto f_h$ is a morphism of Poisson structures:
$$
\{f_{h_1}(x),f_{h_2}(x)\}=\{h_1(\mu),h_2(\mu)\}_{\mathfrak g^*}, \quad \mu=\phi(x).
\leqno(2.5)
$$
Thus, ${\mathcal F}_1$ is closed under the Poisson brackets.
Since $\mathfrak G$ acts in a Hamiltonian way,
${\mathcal F}_2$ is closed under the Poisson brackets as well.

It follows from the Noether theorem that the moment map $\Phi$
is an integral of the Hamiltonian equations for all
Hamiltonian functions $h$ that belong to ${\mathcal F}_2$.
In other words:
$$
\{{\mathcal F}_1,{\mathcal F}_2\}=0.
\leqno(2.6)
$$

\begin{assumption}{\rm
Let a general orbit of the action have dimension $m$.
We shall suppose that
$$
\Span \{ df(x),\; f\in{\mathcal F}_2\}=\ann (T_x(\mathfrak G\cdot x)),
\leqno(2.7)
$$
for general $x\in M$, $\dim \mathfrak G\cdot x = m$.
Whence $\ddim {\mathcal F}_2=2n-m$.
By $\reg{\mathcal F}_2$ denote the open dense set where (2.7) holds.}
\end{assumption}

The assumption 2.1
holds for any proper group action because
all orbits are separated by
invariant functions.
Moreover, for the action of a compact group $\mathfrak G$
the algebra ${\mathcal F}_2$ is generated by a finite number of functions.
To be more precise, let the $\mathfrak G$ action have a finite number
of orbit types. Then there exist functions
$f_1,\dots,f_r\in{\mathcal F}_2$,
such that every function $f\in {\mathcal F}_2$ is of
the form:
$f=F(f_1,\dots,f_r)$.
This theorem was proved by Schwarz [29].
If $M$ is compact then $M$ has a finite number of orbit types.
Furthermore, Mann proved that if $M$ is an orientable manifold
whose homology groups $H_i(M,\Bbb{Z})$ are finitely generated
then the number of orbit types
of any action of a compact Lie group on $M$ is finite. 
A review of results concerning invariant functions of $\mathfrak G$
actions can be found in [27].

The following theorem, although it is a reformulation of
some well known facts about the momentum mapping, is fundamental
in the considerations below.

\begin{theorem}Suppose that assumption 2.1 holds.
Then the algebra of functions ${\mathcal F}_1 + {\mathcal F}_2$ is complete:
$$
\ddim({\mathcal F}_1+ {\mathcal F}_2)+
\dind({\mathcal F}_1+{\mathcal F}_2)=\dim M.
$$
\end{theorem}

{\it Proof.}
We shall need a well known fact that $\ker d\Phi(x)$
is symplectically orthogonal
to the tangent space at $x$ to the orbit of $x$ (see [20]):
$$
\ker d\Phi(x)=(T_x(\mathfrak G\cdot x))^\omega.
\leqno(2.8)
$$

Let $U=\reg({\mathcal F}_1+{\mathcal F}_2)=\reg {\mathcal F}_2$
be the open dense set in $M$, such that for $x\in U$
the moment map (2.1) has maximal rank
(by (2.8) it is equivalent to the fact  that the orbit $\mathfrak G\cdot x$
has maximal dimension) and that (2.7) holds.
Let $x$ belong to $\reg({\mathcal F}_1+{\mathcal F}_2)$.
Consider the linear spaces:
\begin{eqnarray*}
&W_1=W_1(x)=\Span \{\sgrad f(x),\; f\in {\mathcal F}_1\}\subset T_x M \\
&W_2=W_2(x)=\Span \{\sgrad f(x),\; f\in {\mathcal F}_2\}\subset T_x M
\end{eqnarray*}
We shall prove that the symplectic orthogonal complement of $W_1$
coincides with $W_2$:
$$
W_1^\omega=W_2
\leqno(2.9)
$$
Then $(W_1+W_2)^\omega \subset W_1+W_2$ which is equivalent to
the completeness of the algebra
${\mathcal F}_1 + {\mathcal F}_2$ (see remark 1.1).

It remains to prove (2.9)
(note that from (2.6) we have that $W_1$ and $W_2$
are symplectic orthogonal: $\omega(W_1,W_2)=0$).
This condition is equivalent
to the following condition:
$$
\ann(\{df(x), \; f\in {\mathcal F}_1\})=
\ker d\Phi(x)=\Span\{\sgrad f(x),\; f\in {\mathcal F}_2\}.
\leqno(2.10)
$$
On the other side, from (2.7) we get:
$$
\Span \{ \sgrad f(x),\; f\in{\mathcal F}_2\}=(T_x(\mathfrak G\cdot x))^\omega,
$$
which together with (2.8) prove (2.10). The theorem is proved.
q.e.d.

\begin{remark}{\rm 
Note that if $h:\mathfrak g^*\to \Bbb{R}$ is an invariant of the co-adjoint
action then $f_h=h\circ\Phi$ will be a $\mathfrak G$--invariant function on $M$.
Also, we have the following property of functions in ${\mathcal F}_1$.
Let $f_h=h\circ\Phi\in{\mathcal F}_1$ and let $\mu_0=\Phi(x_0)$.
If $ad^*_{dh(\mu_0)}\mu_0$ is equal to zero
(or equivalently if $dh(\mu_0)\in\ann T_{\mu_0}{\mathcal O}(\mu_0)$),
then by (2.4) and (2.5) we
get that $f_h$ commutes with all functions from ${\mathcal F}_2$ at $x_0$.
By the proof of theorem 2.1, 
for $x_0\in\reg({\mathcal F}_1+{\mathcal F}_2)$,
$df_h(x_0)$ belongs to
the $\Span\{ df(x_0),\; f\in{\mathcal F}_2\}$.
Thus we have:
\begin{eqnarray*}
\ddim({\mathcal F}_1+{\mathcal F}_2)&=
&\dim M-\dim \mathfrak G\cdot x + \dim {\mathcal O}(\mu) \\
&=&\dim M +\dim \mathfrak G_x - \dim \mathfrak G_\mu \\
\dind({\mathcal F}_1+{\mathcal F}_2)&=
&\dim \mathfrak G_\mu-\dim \mathfrak G_x,
\end{eqnarray*}
for general $x\in M$, $\mu=\Phi(x)$
($\mathfrak G_\mu$ and $\mathfrak G_x$ denotes the isotropy groups
of $\mathfrak G$ action at $\mu$ and $x$).}
\end{remark}

The following corollary can be derived from theorem 2.1 and the above
observations:

\begin{corollary}
Suppose that assumption 2.1 holds.
Let $\mathcal A$ be any algebra of functions on $\mathfrak g^*$
and ${\mathcal F}_1({\mathcal A})$ be the pull-back of $\mathcal A$ by
the moment map:
$$
{\mathcal F}_1({\mathcal A})=\Phi^*({\mathcal A})=\{f_h=h\circ \phi, \; h\in\mathcal A\}.
$$
Then ${\mathcal F}_1({\mathcal A})+{\mathcal F}_2$ is a complete
algebra on $M$ if and only if
$\mathcal A$ is a complete algebra on the orbit ${\mathcal O}(\mu)$
of the co-adjoint action
of the group $\mathfrak G$ on $\mathfrak g^*$, for general $\mu \in \phi(M)$.
\end{corollary}

Corollary 2.1
 is connected with integrability of the so-called collective motion
(we follow the terminology of Guillemin and Sternberg [18]).
The Hamiltonian $H: M\to \Bbb{R}$ is said to be  {\it collective}
if $H$ is of the form $H=h\circ\Phi$, or in our notation if $H$ belongs
to ${\mathcal F}_1$.

Let a general co-adjoint orbit ${\mathcal O}(\mu)\subset \Phi(M)$
have dimension $2l$.

\begin{theorem}
 Suppose that assumption 2.1 holds.
Let $h:\mathfrak g^*\to\Bbb{R}$ be a Hamiltonian function such that
the Euler equations:
$$
\dot \mu=ad^*_{dh(\mu)}\mu,
$$
are completely integrable on general co-adjoint orbits ${\mathcal O}(\mu)\subset \Phi(M)$ with a set of Lie-Poisson commuting
integrals  $f_1,\dots,f_l: \mathfrak g^*\to\Bbb{R}$.
Then the Hamiltonian equations on $M$ with Hamiltonian function
$H=h\circ\Phi$ are completely integrable.
The complete algebra of first integrals is
$$
\{f_1\circ\Phi,\dots,f_l\circ\Phi\}+{\mathcal F}_2.
$$
\end{theorem}

The above theorem generalizes the results of Guillemin and Sternberg
obtained for the case when
the action $\mathfrak G$ on $M$ is {\it multiplicity-free}
(see [16, 17, 18]).

\begin{remark}{\rm
If $\mathfrak G$ is a compact group, then the connected components
of regular invariant
submanifolds of the integrable systems from theorems 2.1 and 2.2
are isotropic tori of dimension
$\dim \mathfrak G_\mu-\dim \mathfrak G_x$ and
$\frac12(\dim \mathfrak G+\dim \mathfrak G_\mu)-\dim \mathfrak G_x$
respectively.}
\end{remark}

\section{Manifolds all of whose geodesics are closed}

Suppose that an $n$--dimensional Riemannian manifold $(Q,ds^2)$
has the property that  for every $x\in M$,
all geodesics starting from $x$ return back to the same point.
To be more precise, let $\gamma(t)$, $\vert \dot\gamma(t)\vert=1$
be a geodesic line. Then there is  $T\in\Bbb{R}$ such that
$\gamma(0)=\gamma(T)$, $\dot\gamma(0)=\dot\gamma(T)$.
This condition implies, by the theorem of Wadsley [4, 35],
that there is a
common period for all geodesics.
Such Riemannian manifolds are called {\it P--manifolds}.

It is clear that then the geodesic flow
induces an $S^1$-action on $T^*M$, with the moment
map given by the Hamiltonian function.

From theorem 2.1 we obtain:

\begin{theorem}
The geodesic flow on a Riemannian P--manifold is completely integrable.
\end{theorem}

Theorem 3.1 says that periodic trajectories are organized
in $n$--dimensional Lagrangian tori. The frequencies of all tori
are resonant.
More about the geometry of manifolds all of whose geodesics are closed,
together with examples of such manifolds, can be found in [4].

\begin{remark}{\rm
A stronger theorem was stated by Duran [13].  He claimed that commuting
integrals could be taken in such a way that the singular set would be a
polyhedron. But there seem to be some problems in his proof (lemma 3.1
in [13]).}
\end{remark}

\section{Geodesic flows on homogeneous spaces}

An important application of theorems 2.1 and 2.2 is the case when $M$
is the (co)tangent bundle of a
homogeneous space ${\mathfrak G}/\mathfrak H$, where
$\mathfrak G$ is a compact connected Lie group.
In that way we have another proof of results from [8].

Let $\mathfrak g=T_e\mathfrak G$, $\mathfrak h=T_e\mathfrak H$ be
the Lie algebras of $\mathfrak G$ and $\mathfrak H$. Let
$\mathfrak g=\mathfrak h+\mathfrak v$ be the orthogonal
decomposition of $\mathfrak g$ according to a non-degenerate  $Ad_{\mathfrak
G}$-invariant scalar product $\langle\cdot,\cdot\rangle$.  We can identify
$\mathfrak v$ with the tangent space $T_{\pi(e)}(\mathfrak G/\mathfrak H)$, where
$\pi: \mathfrak G \to M$ is the natural projection.  Denote by $g\xi$ the
action of $g\in \mathfrak G$ on the element $\xi \in \mathfrak v=T_{\pi(e)}M$.
All functions will be analytic, polynomial in velocities.

Let $ds_0^2$ be the $\mathfrak G$ invariant metric on $\mathfrak G/\mathfrak H$
defined by:  $\langle g\xi,g\eta\rangle_{\pi(g)}=\langle \xi,\eta
\rangle$.  Identify $\mathfrak g$ and $\mathfrak g^*$ by the scalar product
$\langle\cdot,\cdot\rangle$.

Consider $T(\mathfrak G/\mathfrak H)$ as a symplectic manifold whose symplectic
form is the pull-back of the canonical symplectic form on $T^*(\mathfrak
G/\mathfrak H)$ by the metric $ds_0^2$.  Then the natural $\mathfrak G$ action on
$T(\mathfrak G/\mathfrak H)$ is Hamiltonian with the moment map of the form:
$\Phi(g\xi)=Ad_g\xi$, $\xi\in \mathfrak v$.

There are various constructions
of complete algebras of involutive functions
for compact Lie algebras.
In order to apply theorem 2.2 we shall focus on
the following construction.

Denote by $I(\mathfrak g)$   the algebra of
$Ad_{\mathfrak G}$ invariant polynomials on $\mathfrak g$.
Mishchenko and Fomenko showed that
the polynomials ${\mathcal A}_a=\{p^i_a\}$
obtained from invariant polynomials by shifting the argument:
$$
p(\xi+\lambda a)=\sum p^i_a(\xi)\lambda^i, \quad \xi\in \mathfrak g,
 \quad p \in I(\mathfrak g)
$$
are in involution [22]. Furthermore,
for general regular  $a\in \mathfrak g$, the family ${\mathcal A}_a$ forms a
complete involutive set of functions on
{\it every} adjoint orbit in
$\mathfrak g$.
For regular orbits this is proved by Mishchenko and Fomenko [22].
For singular orbits there are several different proofs:
by Mikityuk [24], Brailov [10] and Bolsinov [5].

Since $a$ is regular, $\mathfrak g_a=\{\eta\in \mathfrak g,\; [\eta,a]=0\}$
is a Cartan subalgebra. Let $b$ belong to $\mathfrak g_a$ and let
$D: \mathfrak g_a\to \mathfrak g_a$
be a symmetric operator. By $\varphi_{a,b,D}$ denote
the symmetric operator (called {\it sectional operator})
defined according to the orthogonal decomposition:
$\mathfrak g=\mathfrak g_a+[a,\mathfrak g]$:
$$
\varphi_{a,b,D}\vert_{\mathfrak g_a}=D,
\quad \varphi_{a,b,D}\vert_{[a,\mathfrak g]}=ad_a^{-1}ad_b.
$$
The functions ${\mathcal A}_a$ are integrals of the Euler equations
(for example see [33]):
$$
\dot \xi=[\xi,\nabla h_{a,b,D}(\xi)],\quad
h_{a,b,D}(\xi)=\frac12\langle \varphi_{a,b,D}(\xi),\xi \rangle.
$$

In the case of compact Lie groups, among the sectional operators there are
positive definite ones.  Then
$H_{a,b,D}=h_{a,b,D}\circ\Phi$ is the Hamiltonian of the geodesic flow of
a certain metric that we shall denote by $ds_{a,b,D}^2$.

From theorem 2.1 we obtain the following theorem:

\begin{theorem} {\bf [8]}
Let $Q$ be a homogeneous space ${\mathfrak G}/\mathfrak H$, where
$\mathfrak G$ is a compact connected Lie group.
Then the geodesic flows of the metrics $ds^2_{a,b,D}$ on $Q$ are
completely integrable.
In particular, for $\varphi_{a,b,D}=Id_\mathfrak g$ we have integrability
of the geodesic flow of the  metric $ds^2_0$.
\end{theorem}

Previous results include the case of integrability
of geodesic flows on compact Lie groups [22], symmetric spaces
[10, 21, 24, 32] and the spaces $\mathfrak G/\mathfrak H$
where $(\mathfrak G,\mathfrak H)$ form a
Gelfand (or spherical) pair [16, 25]
(exception are $SO(n)/SO(n-2)$ [32] and $SU(3)/T^2$ [28]).
For symmetric spaces and spherical pairs, 
in a neighborhood of a generic point $x\in T(\mathfrak G/\mathfrak H)$ each 
$\mathfrak G$--invariant function $f$ can be expressed as 
$f=h \circ \Phi$ and 
thus we can use just functions from $\mathcal F_1$ to get the integrability of
any $\mathfrak G$--invariant geodesic flow on $\mathfrak G/\mathfrak H$.
Spherical pairs are classified in [19, 25].

Examples given in [6, 7, 11] are homogeneous spaces $\mathfrak G/\Gamma$
of noncompact groups,
where $\Gamma\subset \mathfrak G$ is a discrete cocompact subgroup. The corresponding  geodesic flows are integrable by smooth integrals,
and by Taimanov's theorem [30, 31] can not be integrable by analytic ones.

\section{Bi-quotients of Lie groups}

Let $\mathfrak G$ be a compact connected Lie group.
Consider a subgroup $\mathfrak U$ of $\mathfrak G\times\mathfrak G$ and
define the action of $\mathfrak U$ on $\mathfrak G$ by:
$$
(g_1,g_2)\cdot g =g_1gg_2^{-1}, \quad (g_1,g_2)\in \mathfrak U,\; g\in\mathfrak G.
$$
If the action is free then the orbit space $\mathfrak G/\mathfrak U$ is a
smooth manifold called a
{\it bi-quotient} of the Lie group $\mathfrak G$.
In particular, if $\mathfrak U=\mathfrak K\times\mathfrak H$, where
$\mathfrak K$ and $\mathfrak H$ are subgroups of $\mathfrak G$, then
the bi-quotient of $\mathfrak G$
is denoted by ${\mathfrak K}\backslash {\mathfrak G}/{\mathfrak H}$.
The condition that $\mathfrak U=\mathfrak K\times\mathfrak H$ acts freely on $\mathfrak G$
is:
$$
g{\mathfrak K}g^{-1} \cap \mathfrak H = e, \quad
\mathrm{for\; any}\ \ g\in
\mathfrak G.
$$

The bi-invariant metric on $\mathfrak G$ gives the Riemannian metric
on ${\mathfrak K}\backslash {\mathfrak G}/{\mathfrak H}$.
Denote such metric by $ds_0^2$.

In this section we shall prove the following general statement:

\begin{theorem}
Let $Q$ be a bi-quotient ${\mathfrak K}\backslash {\mathfrak G}/{\mathfrak H}$, where
$\mathfrak G$ is a compact connected Lie group.
Then the geodesic flow of the metric $ds^2_0$ on $Q$
induced by a bi-invariant metric on $\mathfrak G$ is
completely integrable in the non-commutative sence by means of
analytic integrals, and therefore in the classical commutative sence
by means of $C^\infty$--smooth integrals.
\end{theorem}

{\it Proof.}
We shall use the following description of tangent spaces of
${\mathfrak K}\backslash {\mathfrak G}/{\mathfrak H}$. By definition the elements of
this bi-quotient space are equivalence classes of the form
$\{kgh\}$ where $g$ is fixed, and $k$ and $h$ run over ${\mathfrak
K}$ and ${\mathfrak H}$ respectively.

The tangent space for this equivalence class at the point $g$ can be
represented in the form
$$
g\mathfrak h + \mathfrak k g,
\leqno(5.1)$$
so that the tangent space to the bi-quotient at the point $\mathfrak K g
\mathfrak H$ can naturally be represented as the orthogonal complement
$(g\mathfrak h + \mathfrak k g)^\perp$ with respect to the bi-invariant metric on
$\mathfrak G$.  If we want this definition not to depend on the choice of a
representative in the class, we must explain how these orthogonal
complements are identified at points $g$ and $kgh$. Since $kgh$ is
represented in this form uniquely, the elements $k$ and $h$ are well
defined. Then we can consider the map $L_k \circ R_h$ which maps $g$ onto
$kgh$ and take its differential. As a result, the subspace $g\mathfrak h +
\mathfrak k g$ is mapped onto $kg\mathfrak h h + k\mathfrak k gh$, which obviously
coincides with $kgh\mathfrak h + \mathfrak k kgh$.  The orthogonal complements
will also be matched.  Such an identification is natural because the
corresponding vectors from the orthogonal complements are projected onto
the same vector tangent to the bi-quotient.

The next thing we want to do is to describe some important functions on
$T(\mathfrak K\backslash \mathfrak G/\mathfrak H)$ using their liftings onto 
$T\mathfrak G$.  
In what follows, we shall denote by ${\mathcal O}_\mathfrak K(\xi)$ and
${\mathcal O}_\mathfrak H(\xi)$ the orbits through $\xi\in \mathfrak g$ of the natural
adjoint actions of subgroups $\mathfrak K$ and $\mathfrak H$ respectively.
All functions will be analytic, polynomial in velocities.

Let $f$ be a function on $T\mathfrak G$ which is right-invariant
(with respect to $\mathfrak G$) and left-invariant with respect to $\mathfrak K$
(equivalently, $f$ can be seen as a function obtained
by right-translations from some
$Ad_\mathfrak K$ invariant function on $\mathfrak g$).
Then this function being restricted onto the orthogonal complement is well projected onto $T(\mathfrak K\backslash \mathfrak G/\mathfrak H)$.
Denote the space of functions so obtained by $\mathcal F_1$.

Analogously, let $f$ be a function on $T\mathfrak G$ which is left-invariant (with respect to $\mathfrak G$) and right-invariant with respect to $\mathfrak H$.
Then this function being restricted onto the orthogonal complement is well projected onto $T(\mathfrak K\backslash \mathfrak G/\mathfrak H)$.
Denote the space of functions so obtained by ${\mathcal F}_2$.
Note that $\{{\mathcal F}_1,{\mathcal F}_2\}=0$.

These functions give the family of first integrals $\mathcal F_1 + {\mathcal F}_2$
for the geodesic flow of the metric $ds_0^2$ on
$\mathfrak K \backslash\mathfrak G /\mathfrak H$.

The problem is to verify the completeness of this family.

The functions we have constructed  have 
another natural description.
Consider the following subset in $\mathfrak g$:
$$
C=\mathfrak k^\perp \cap \{g\mathfrak h^\perp g^{-1},\; g\in \mathfrak G\}=
\cup_{g\in\mathfrak G}
(\mathfrak k+g\mathfrak h g^{-1})^\perp
\leqno(5.2)$$
This set is ${Ad}_\mathfrak K$-invariant. So we can consider $\mathfrak K$--invariant 
functions on it.  Take such a function and extend it by right
translations to the whole of $T\mathfrak G$. Actually we shall obtain a
function defined for each fiber $T_g\mathfrak G$ only on the subset $Cg$, but
this subset contains $(\mathfrak k g + g\mathfrak h)^\perp$ because $(\mathfrak k g +
g\mathfrak h)^\perp=(\mathfrak k+g\mathfrak h g^{-1})^\perp g \subset Cg$.  We should
remark that $C$ is not smooth in general.  But being an algebraic set it
is smooth almost everywhere.

By ${\mathcal F}_C$ denote the set of $Ad_\mathfrak K$ invariant functions on $C$ and
by $\reg C$ the set of regular orbits
${\mathcal O}_\mathfrak K(\xi)$ of $Ad_\mathfrak K$ action on $C$.

From the choice of the functions ${\mathcal F}_C$, we have the important relation
that the number of functions in ${\mathcal F}_C$ is equal to the number
of functions in $\mathcal F_1$:
$$
\ddim{\mathcal F}_C=\ddim{\mathcal F_1}
\leqno(5.3)
$$

The number of independent $\mathfrak K$--invariant functions on $C$ is equal
to the difference:
$$
\ddim{\mathcal F}_C=\dim C - \dim {\mathcal O}_\mathfrak K
(\xi),\quad \xi\in\reg C
\leqno(5.4)
$$

On the other hand, the tangent
space of $C$ in $\xi$ can be represented as the intersection of 
$T_\xi\{g\mathfrak h^\perp g^{-1},\; g\in \mathfrak G\}$  and
$\mathfrak k^\perp$.  
The tangent plane of the set $\{g\mathfrak h^\perp g^{-1},\;
g\in \mathfrak G\}$ at a generic point $\xi$ can be obviously  presented as
$\mathfrak h^\perp + [\xi,\mathfrak g]$. Thus:
$$
T_\xi C = (\mathfrak h^\perp + [\xi,\mathfrak g])\cap \mathfrak k^\perp.
\leqno(5.5)
$$
Also notice that $\dim  {\mathcal O}_\mathfrak K (\xi)=\dim [\xi,\mathfrak k]$.
Thus, by (5.5) the number of $\mathfrak K$-invariant functions is equal to
$$
\dim (\mathfrak h^\perp + [\xi,\mathfrak g])\cap \mathfrak k^\perp - \dim [\xi,\mathfrak k].
\leqno(5.6)
$$

Let us compute this number.
We have
\begin{eqnarray*}
&&\dim (\mathfrak h^\perp + [\xi,\mathfrak g])\cap \mathfrak k^\perp=\\
&&\dim(\mathfrak h^\perp + [\xi,\mathfrak g])+\dim\mathfrak k^\perp-\dim
(\mathfrak h^\perp + [\xi,\mathfrak g] + \mathfrak k^\perp)=\\
&&(\mathrm{we \;use \;that \;}\mathfrak h^\perp + \mathfrak k^\perp=
\mathfrak g)\\
&&\dim (\mathfrak h^\perp + [\xi,\mathfrak g]) - \dim \mathfrak k 
\end{eqnarray*}

Next,
\begin{eqnarray*}
&&\dim [\xi,\mathfrak k]=\dim \mathfrak k - \dim \mathfrak k_\xi =\\
&&\dim \mathfrak k - \dim (\mathfrak g_\xi \cap \mathfrak k) =\\
&&\dim \mathfrak k - \dim ([\xi,\mathfrak g]^\perp \cap \mathfrak k)=\\
&&\dim \mathfrak k - \dim \mathfrak g + \dim([\xi,\mathfrak g]+\mathfrak k^\perp),
\end{eqnarray*}
where $\mathfrak k_\xi=\{\eta\in\mathfrak k,\; [\eta,\xi]=0\}$,
$\mathfrak g_\xi=\{\eta\in\mathfrak g,\; [\eta,\xi]=0\}$.

Thus, by (5.3) and (5.6) we get:
\begin{eqnarray*}
\ddim{\mathcal F}_1 &=&
\dim (\mathfrak h^\perp + [\xi,\mathfrak g]) - \dim \mathfrak k-\dim \mathfrak k +
\dim \mathfrak g - \dim([\xi,\mathfrak g]+\mathfrak k^\perp)=\\
&=&\dim \mathfrak g- 2\dim \mathfrak k + \dim (\mathfrak h^\perp + [\xi,\mathfrak g]) -  \dim([\xi,\mathfrak g]+\mathfrak k^\perp),
\end{eqnarray*}
for $\xi$ that belongs to the open dense subset $\reg C \subset C$.

Now one can notice that the result for the number of  functions
in the family ${\mathcal F}_2$
will be just the same (after interchanging $\mathfrak k$ and $\mathfrak h$).

Finally, we see that
\begin{eqnarray*}
&&\ddim {\mathcal F}_1 + \ddim {\mathcal F}_2 = \\
&&\dim \mathfrak g - 2\dim \mathfrak k +
\dim (\mathfrak h^\perp + [\xi,\mathfrak g]) -  \dim([\xi,\mathfrak g]+\mathfrak k^\perp)+\\
&&\dim \mathfrak g - 2\dim \mathfrak h + \dim (\mathfrak k^\perp + [\xi,\mathfrak g]) -
\dim([\xi,\mathfrak g]+\mathfrak h^\perp)=\\
&&2(\dim \mathfrak g - \dim \mathfrak k - \dim \mathfrak h)=
2\dim \mathfrak K\backslash \mathfrak G/\mathfrak H.
\end{eqnarray*}

The theorem follows from the lemma below.
q.e.d.

\begin{lemma}
Let $\mathcal F_1$ and ${\mathcal F}_2$ be two Poisson subalgebras 
of $(C^\infty(M),\{\cdot,\cdot\}_M)$ on a symplectic
manifold $M$ that commute, i.e., $\{{\mathcal F}_1,{\mathcal F}_2\}=0$.
If 
$$
\ddim {\mathcal F}_1 + \ddim {\mathcal F}_2 = \dim M,
$$
then ${\mathcal F}_1+{\mathcal F}_2$ is a complete algebra of functions on $M$.
\end{lemma}

{\it Proof.}
Let $x$ belong to $\reg{\mathcal F}_1 \cap \reg{\mathcal F}_2$.
Consider linear spaces:
\begin{eqnarray*}
&W_1=W_1(x)=\Span \{\sgrad f(x),\; f\in {\mathcal F}_1\}\subset T_x M,\\
&W_2=W_2(x)=\Span \{\sgrad f(x),\; f\in {\mathcal F}_2\}\subset T_x M.
\end{eqnarray*}
Under the suppositions of the lemma, we have:
$$
\dim W_1+\dim W_2=\dim M, \quad
\omega(W_1,W_2)=0.
\leqno(5.7)
$$
From $\omega(W_1,W_2)=0$ it follows that
symplectic orthogonal complement of $W_1+W_2$ contains $W_1 \cap W_2$:
$W_1\cap W_2 \subset (W_1+W_2)^\omega$.
On the other hand,  (5.7) implies:
$$
\dim (W_1\cap W_2)=\dim (W_1+W_2)^\omega= \dim M - \dim (W_1+W_2)
$$
Therefore, $W_1\cap W_2=(W_1+W_2)^\omega\subset W_1+W_2$ and
by remark 1.1,  ${\mathcal F}_1+{\mathcal F}_2$ is a complete algebra of functions on $M$.
q.e.d.

\end{document}